\documentclass[manuscript,nonacm]{acmart}

\usepackage{booktabs}
\usepackage{balance}

\begin{document}

\title{Who Is in the Room? Stakeholder Perspectives on AI Recording in Pediatric Emergency Care}
\subtitle{Accepted at the ACM Interactive Health Conference (IH~'26), Porto, Portugal, July 2026.\\DOI: \url{https://doi.org/10.1145/3786579.3804950}}

\author{Alexandre De Masi}
\affiliation{%
  \institution{University of Geneva}
  \city{Geneva}
  \country{Switzerland}
}
\email{alexandre.demasi@unige.ch}

\author{Sergio Manzano}
\affiliation{%
  \institution{University Hospitals of Geneva}
  \city{Geneva}
  \country{Switzerland}
}
\email{sergio.manzano@hug.ch}

\author{Johan N. Siebert}
\affiliation{%
  \institution{University Hospitals of Geneva}
  \city{Geneva}
  \country{Switzerland}
}
\email{johan.siebert@hug.ch}

\author{Fr\'ed\'eric Ehrler}
\affiliation{%
  \institution{University Hospitals of Geneva}
  \city{Geneva}
  \country{Switzerland}
}
\email{frederic.ehrler@hug.ch}

\begin{abstract}
Artificial intelligence systems that record voice and video during pediatric emergencies are emerging as human-computer interaction (HCI) technologies with direct implications for clinical work, promising improvements in documentation, team performance, and post-event debriefing. Yet the perspectives of those most affected, including clinicians, parents, and child patients, remain largely absent from the design and governance of these technologies. This position paper argues that this has direct consequences for the legitimacy and effectiveness of these systems. We examine four areas where these missing perspectives prove consequential (consent, emotional impact, surveillance dynamics, and participatory governance) and propose four positions for reorienting AI recording in pediatric emergency care toward stakeholder-centered HCI inquiry.
\end{abstract}


\maketitle

\section{Introduction}

In a growing number of hospitals, cameras and microphones now capture pediatric resuscitations for artificial intelligence systems designed to document procedures, assess team performance, and feed machine-learning models intended to improve future outcomes~\cite{nolan2020trauma, li2021video, zhang2024focusing}. In emergency medical services, body-worn cameras record paramedic-led resuscitations for quality review and education~\cite{dewar2019bodyworn}. These systems are advancing quickly. Computer vision can recognize concurrent trauma activities at the bedside~\cite{li2021video}. Deep-learning models can detect patient mobilization activities in the intensive care unit (ICU)~\cite{yeung2019computer}. Ambient AI scribes can transcribe clinical encounters in real time~\cite{tierney2024ambient}. Video-based models can predict hospitalization from brief emergency department clips~\cite{omiye2024hospitalization}.

The clinical rationale for these technologies is well-supported. Video-based debriefing improves neonatal resuscitation quality and guideline adherence~\cite{skare2018video}. Collaborative cognitive aids that integrate visual and temporal information can enhance team coordination during acute care~\cite{gonzales2016visual}. Recording and reviewing resuscitations supports learning, self-reflection, and quality improvement~\cite{hill2022learning, heesters2022videoreflection}. The evidence base for what these systems can achieve is growing steadily.

What has not kept pace is sustained attention to the people these systems affect and how they experience being recorded, analyzed, and made legible to algorithms. Recent work on human-centered AI in healthcare identifies a persistent gap between the technical maturity of AI systems and the human-centered research needed to govern their deployment~\cite{andersen2023humancentered}. We argue that this gap is especially consequential in pediatric emergency recording, where the conversations shaping design and governance have included primarily those who build and deploy these systems (engineers, researchers, and institutions) while those whose bodies, voices, grief, decisions, and professional identities are captured (clinicians, parents, and patients) remain largely outside the room. This absence from the design and governance of recording systems has ethical, clinical, and epistemic consequences that warrant closer examination.

From an HCI perspective, this is a design problem as much as an ethical one. The human--computer interaction community has long recognized that technologies are not neutral instruments but sociotechnical systems whose design embeds values, configures relationships, and distributes power among stakeholders~\cite{cruzmartinez2021vsd}. Recording systems in pediatric emergencies are no exception. Decisions about what is captured, who controls the footage, how it is processed, and who can access it are interaction design decisions with direct consequences for the people involved. We approach these questions from within an ongoing empirical research program involving clinicians, families, and institutional stakeholders in pediatric emergency settings, from which we draw both our orientation toward stakeholder-centered inquiry and our understanding of the practical constraints that shape recording practices in these environments.

In what follows, we examine four areas where the absence of stakeholder perspectives undermines the foundations of AI-enabled recording in pediatric emergencies, and argue that addressing them requires a reorientation from performance-centered to stakeholder-centered design. We conclude with four positions intended for the research and practice communities working at the intersection of human-centered computing and health, each directed toward ensuring that those most affected by these systems have a place in the conversations that shape them. Our aim is not to slow the adoption of these technologies but to strengthen their long-term legitimacy and sustainability by grounding them in the perspectives of the people they affect.

\section{Approach and Scope}

This is a position paper, not a systematic review or an empirical study. We develop our argument through a structured synthesis of literature spanning HCI, clinical ethics, surveillance studies, and pediatric emergency medicine, drawing connections across bodies of work that have largely developed in isolation. While existing work in health AI ethics has identified broad concerns around fairness, transparency, and accountability, and value-sensitive design research has proposed general methods for stakeholder engagement, neither tradition has examined the specific intersection of AI-enabled audiovisual recording and pediatric emergency care, where consent, emotional intensity, surveillance dynamics, and child vulnerability converge simultaneously. Our orientation toward these questions is informed by an ongoing empirical research program in which we work with clinicians, families, and institutional stakeholders in pediatric emergency settings to design and evaluate interactive clinical decision support tools. This program provides both our understanding of the practical constraints that shape recording practices and our commitment to stakeholder-centered inquiry as a precondition for legitimate technology deployment.

\section{Missing Perspectives}

\subsection{Consent}

Informed consent is a cornerstone of both clinical ethics and technology governance. Yet in pediatric emergencies, the conditions that make consent meaningful, including time for reflection, comprehension, emotional capacity, and genuine voluntariness, are precisely what the crisis diminishes~\cite{denboer2018ethical}. In practice, the urgency of resuscitation rarely allows clinicians to seek full informed consent at the outset, and discussion is often necessarily deferred until the child is stabilized. Parents of a child undergoing resuscitation are in acute psychological distress, and clinicians are focused on keeping a child alive. Under these conditions, meaningful consent to AI-mediated audiovisual recording is difficult to obtain in any substantive sense~\cite{iserson2024informed}.

Existing frameworks offer partial but insufficient responses to this tension. Practical guidance on informed consent for AI in emergency medicine acknowledges that in life-threatening circumstances consent may need to be \emph{presumed}~\cite{iserson2024informed}. Tiered frameworks for evaluating when patients should be consented about AI were developed for healthcare broadly, not for the particular affective intensity of pediatric settings~\cite{roberts2024ethically}. Deferred consent is most acceptable during low-risk interventions with narrow therapeutic windows~\cite{woolfall2022deferred}, conditions that do not readily map onto continuous, identity-capturing audiovisual recording. In neonatal care specifically, informed consent before delivery-room recording is often impractical, and ethical requirements diverge across quality assurance, research, and educational uses of the same footage~\cite{denboer2018ethical}. A single recording may simultaneously serve documentation, debriefing, training, algorithmic development, and institutional quality metrics, yet consent obtained for one purpose does not extend to another. Current consent architectures are ill-equipped for this kind of purpose multiplicity~\cite{gerke2020ethical}.

From an interaction design standpoint, this points to a fundamental mismatch between the temporal structure of consent interfaces, which typically operate as point-in-time agreements, and the temporally distributed, multi-purpose nature of clinical recording. At the governance level, international guidance~\cite{who2021ethics} and recent commentary~\cite{mello2025regulation} both acknowledge that governance infrastructure has not kept pace with AI adoption, and emerging frameworks for governing health AI data emphasize the need for mechanisms that can accommodate evolving consent requirements over time~\cite{morley2022governing}. Proposals for pre-established institutional norms and post-hoc notification when real-time consent is infeasible~\cite{vayena2018systemic} are procedurally reasonable, but they address institutional process rather than the relational trust between families, clinicians, and the technologies present in their shared clinical space. Designing consent as an ongoing interaction rather than a one-time event requires moving beyond interface-level solutions toward the kind of sustained, context-sensitive inquiry into stakeholder needs that pediatric emergency care demands.

\subsection{Recording as Emotional Intervention}

Pediatric emergencies are among the most emotionally intense events in clinical medicine, and there is growing evidence that recording these events does not merely document them but alters them in significant ways. In pediatric trauma care, qualitative research on trauma video review has shown that recording shapes decision-making, constrains emotional expression, and reconfigures accountability structures among team members~\cite{dainty2021staffperceptions}. These findings suggest that the recording device should be understood not as a passive data-collection instrument but as an active participant in the sociotechnical configuration of the clinical encounter.

The emotional consequences appear to differ substantially across stakeholder groups. For parents, some value live video as reassurance about their child's care, while the same research identifies risks of anxiety, re-traumatization, and perceived loss of control~\cite{caeymaex2020parents}. Sharing resuscitation footage with parents can support understanding, but requires careful mediation by clinicians with relational sensitivity~\cite{denboer2021reviewing}. The distinction between offering footage as a resource and recording without meaningful consultation raises important questions about relational ethics in design. For clinicians, the presence of recording introduces what might be understood as a dual audience: the patient and family in front of them, and an invisible institutional observer behind the lens. Staff generally accept recordings but emphasize the importance of data security and emotional protection~\cite{glancova2021videorecognition}. Providers recognize benefits of video review but simultaneously express concern about surveillance and post-hoc judgment of decisions made under extreme time pressure~\cite{denboer2019benefits}. Survey evidence indicates that approximately half of physicians oppose recording clinical encounters, citing fears of litigation~\cite{tsulukidze2020attitudes}.

For child patients themselves, the long-term emotional implications remain largely unexamined. Recordings made during pediatric resuscitations capture children at their most vulnerable, and this footage may persist in institutional systems long after the event. As these children grow, the possibility of encountering their own resuscitation footage, or knowing that it exists and may have been used for algorithmic training, raises questions about autonomy, dignity, and psychological safety that current design frameworks do not address.

These findings indicate that recording alters the emotional and relational conditions under which care is delivered. Clinical practice guidelines for family-centered care in the ICU emphasize communication, shared decision-making, and family presence as core values~\cite{davidson2017guidelines}. These guidelines describe a set of design requirements that AI recording systems should be evaluated against, yet current system evaluations focus almost exclusively on technical performance. AI recording systems designed without attention to these relational dimensions risk working against the care practices they are intended to support.

\subsection{The Surveillance Effect}

Where the previous section examined how recording alters the emotional and relational experience of care, we turn here to a distinct but related concern: the systematic behavioral effects of being observed and their epistemological consequences for the data these systems produce. There is consistent evidence that observation changes behavior in clinical settings. Systematic reviews have established that research participation itself influences behavior~\cite{mccambridge2014hawthorne}, and a meta-analysis of video camera observation confirmed that camera presence produces measurable self-reported behavioral change, though at lower rates than direct human observation~\cite{demetriou2025evaluating}. In hand hygiene compliance, for instance, the effect has been measured at 3.75 events per hour when clinicians were observed versus 1.48 when they were not~\cite{hagel2015hawthorne}. Studies of physician behavior under direct observation report that those who acknowledge a behavioral effect describe specific and clinically meaningful changes in their practice~\cite{goodwin2017hawthorne}.

These findings raise an important epistemological question for AI recording in clinical care: whether the system captures authentic practice or performative compliance. If clinicians modify their emotional expression, avoid unconventional-but-appropriate clinical decisions, or prioritize documentable actions over clinical judgment when aware of recording, the data these AI systems generate may be systematically biased, and models trained on such data would encode performance rather than practice. This concern is compounded by technical limitations. Speech recognition for clinical documentation shows significant error rates requiring manual correction~\cite{blackley2019speech}, and analyses of the digital scribe pipeline identify a gap between ambient AI scribe capability and clinical deployment readiness~\cite{quiroz2019digital}. When the recording infrastructure is both behaviorally distorting and technically imperfect, the epistemic foundations of AI models built on that data warrant scrutiny.

Theoretical frameworks from surveillance studies and HCI help to foreground the power dynamics involved. Critical analysis of health surveillance technologies suggests that emergency and safety framing facilitates acceptance of monitoring that may subsequently become entrenched~\cite{couch2020panopticon}. Work on algorithmic control in the workplace identifies six mechanisms through which algorithms govern workers, including recording, rating, and restricting, creating conditions associated with disempowerment and chronic stress~\cite{kellogg2020algorithms}. The concept of ``sousveillance'' raises the question of whose gaze is privileged when recording technologies are deployed in asymmetric settings: whether the institution watches the worker, or whether those being observed can reclaim meaningful agency over the recording process~\cite{mann2003sousveillance}. Research on video surveillance in psychiatric units suggests that evidence does not support surveillance as an effective safety mechanism in shared clinical spaces, and recommends case-by-case ethical evaluation rather than routine deployment~\cite{appenzeller2020video}.

Importantly, work on video-reflexive ethnography demonstrates that the same recording technology can function either as imposed top-down surveillance or as a participatory reflexive tool, depending on how it is deployed and who controls the process~\cite{iedema2011reflexivity, heesters2022videoreflection}. This distinction between surveillance and reflexivity is central to the question of ethical legitimacy in AI recording for pediatric emergency care. It underscores that the ethical character of a recording system is not a property of the technology itself but of the interaction design decisions that govern its deployment, access, and use.

\subsection{Participation in Design}

Given the challenges outlined above, participatory approaches to design might seem a natural response. The HCI community has developed participatory design and value-sensitive design (VSD) methods for health technology over several decades~\cite{cruzmartinez2021vsd, andersen2023humancentered}. Yet their application to AI in healthcare remains remarkably limited. The first narrative review at the intersection of VSD, AI, and healthcare screened 819 articles and identified only nine that met criteria for meaningful engagement with value-sensitive design~\cite{long2024vsd}. Most of those nine addressed only individual-level values such as trust and autonomy, while organizational and societal values, including equity, justice, and institutional accountability, were largely absent. A systematic review of human-centered AI clinical decision support documented persistent workflow misalignment arising in part because designers did not engage clinicians iteratively throughout the development lifecycle~\cite{wang2023humancentered}. A scoping review of AI scribes found that studies assessed almost exclusively technical validity, with few examining clinical usability or the experience of being transcribed~\cite{vanbuchem2021scribe}. Broader work on AI ethics in healthcare has argued that meaningful stakeholder engagement, including the participation of non-technical staff, patients, and families, is a precondition for legitimate AI governance, not an optional supplement~\cite{kaye2024stakeholder}.

Where participatory methods have been applied in adjacent healthcare domains, results demonstrate their value. Co-design with emergency department patients and caregivers has revealed needs that technical approaches alone are unlikely to surface, including the desire for dynamic, real-time information and active participation in diagnostic processes~\cite{seo2025designing}. Human-centered design with parents and triage nurses for an AI-powered pediatric ED tool indicated that involving families can increase parents' sense of agency and control~\cite{litwin2025parent}, precisely the values that non-consultative recording can erode. Co-design of explainable AI with clinicians, including doctors, nurses, and paramedics, has shown that explanation interfaces matter as much as the underlying algorithm~\cite{panigutti2023codesign}. Iterative co-design with clinicians has also revealed that current explainability approaches are poorly suited to time-constrained clinical environments~\cite{jacobs2021designing}, the very settings where AI recording systems are proliferating fastest~\cite{tierney2024ambient, zhang2024focusing}.

In the pediatric domain, the argument for participatory governance is strengthened by the fact that the population served cannot advocate for itself. Children's rights are frequently treated as secondary in digital health design~\cite{holly2020children}. Ethical AI in pediatrics calls for diverse implementation teams and careful attention to the complexity of pediatric data, which involves multiple informants with potentially divergent interests~\cite{boch2022ethical, who2021ethics}. The PEARL-AI framework represents a first step toward child-centered ethical governance of medical AI, addressing harm prevention, autonomy, consent, and data protection across all phases of the AI lifecycle~\cite{chng2025pearl}. The tools for meaningful stakeholder engagement exist, and evidence indicates that stakeholders are willing to participate~\cite{caeymaex2020parents, seo2025designing, denboer2019benefits}. What is lacking is the institutional and design culture that would prioritize such consultation.

\section{Positions for Design and Governance}

The concerns raised above are deeply entangled and reveal a flawed assumption: that technical merit is sufficient to legitimate the deployment of AI recording systems. We propose four positions that follow from the evidence and arguments presented, framed as contributions to ongoing discussions within the HCI and health informatics communities about the responsible design of AI in clinical settings. We recognize that each position involves substantial organizational complexity, and we advance them not as prerequisites for deployment but as trajectories along which recording systems can progressively evolve toward greater stakeholder legitimacy.

\paragraph{Position 1: Consent for AI recording in pediatric emergencies should be designed as an ongoing, purpose-specific governance process rather than a point-in-time agreement.}
A single recording in a pediatric resuscitation may serve documentation, debriefing, training, algorithmic development, and quality review simultaneously~\cite{denboer2018ethical, gerke2020ethical}, and each purpose carries distinct ethical weight. We argue that consent architectures for these systems should be purpose-specific, temporally distributed, and revisable, enabling families and clinicians to grant or withdraw consent for specific downstream uses after the emergency has passed, with default restrictions on use until affirmative consent is obtained. In practice, this could mean that recording begins under institutional authority during the emergency, but within 48 hours a structured digital interface contacts the family, explains each potential use of the footage (clinical debrief, team training, algorithmic development), and allows them to authorize or restrict each use independently, with defaults set to the most restrictive option. This would require designing data governance infrastructure alongside data capture infrastructure, an area where HCI methods for iterative, user-centered interface design could make a direct contribution. Implementing such infrastructure is technically challenging. Hospital information technology (IT) ecosystems are often characterized by fragmented electronic health records with no standard data provenance across documentation, training, and quality assurance silos. This fragmentation is itself a design research opportunity rather than a reason to defer, and understanding it should inform the consent architectures we advocate.

\paragraph{Position 2: The emotional and relational impact of recording should be evaluated and reported as a core outcome in system evaluations, alongside technical performance.}
Evidence indicates that recording transforms emotional expression, clinical decision-making, and parent-clinician relationships~\cite{dainty2021staffperceptions, caeymaex2020parents, denboer2019benefits}, yet no standard evaluation framework for AI recording systems in emergency care includes emotional or relational impact as a reported outcome. System evaluations should include measures of clinician-reported stress and perceived surveillance, parental experience of recording including perceived agency and distress, and observable changes in team communication. Well-established methods exist for evaluating the experiential and relational dimensions of interactive systems. Extending them to AI recording evaluations would help ensure that performance claims are not made in isolation from the human consequences of deployment. In practice, this could involve validated instruments for clinician workload and perceived surveillance, structured post-event interviews with families about their experience of being recorded, and interaction analysis of team communication patterns under recording and non-recording conditions. These methods are already used in HCI evaluation research; what is needed is their systematic application to AI recording contexts.

\paragraph{Position 3: AI recording systems should default to a reflexive model rather than a surveillance model, with the distinction made explicit in system design.}
Evidence indicates that the same recording technology produces substantially different outcomes depending on whether it is deployed as institutional surveillance or as a clinician-controlled reflexive practice tool~\cite{iedema2011reflexivity, heesters2022videoreflection}. We propose that AI recording systems in pediatric emergency care should default to clinician-controlled, team-reviewed footage that is not automatically available for institutional performance metrics, algorithmic training, or third-party analysis unless separate, purpose-specific consent is obtained. This position does not oppose institutional learning or AI development; rather, it argues that the power dynamics embedded in recording architecture should be made visible and subject to negotiation. Designing for reflexivity rather than surveillance is fundamentally an interaction design choice, one that shapes who has agency within the system. Importantly, reflexive defaults must be paired with family access pathways. Evidence indicates that parents value access to recordings for understanding their child's care and processing the experience~\cite{caeymaex2020parents, denboer2021reviewing}. The goal of clinician-controlled defaults is to protect against institutional surveillance, not to limit family transparency.

\paragraph{Position 4: Governance of pediatric emergency AI recording systems should recognize children as rights-holders, not merely data subjects.}
Consider a three-year-old recorded during a resuscitation: at eighteen, they may discover that footage of their most vulnerable moment has been used to train commercial AI systems, without their knowledge or any mechanism for them to have objected. Children cannot consent for themselves, their data is generated during events of extreme vulnerability, and recordings capture not only clinical procedures but the emotional responses of families in crisis. Child-centered frameworks such as PEARL-AI~\cite{chng2025pearl} and rights-based approaches grounded in the UN Convention on the Rights of the Child~\cite{holly2020children} offer useful starting points for governance that takes children's interests seriously from the design phase onward. Governance bodies for pediatric AI recording should include parent representatives, pediatric ethics expertise, and, where appropriate, youth advocates, and that data retention policies should account for the fact that recordings of children can have consequences extending across their lifetime. Precedents in other domains offer design patterns for age-differentiated governance. The General Data Protection Regulation (GDPR), in Article 8, establishes age-of-consent provisions for children's data in information society services~\cite{gdpr2016}, and the US Children's Online Privacy Protection Act imposes specific obligations for data collected from minors~\cite{coppa1998}. Health recording raises analogous but more acute challenges given the sensitivity of clinical footage and the impossibility of meaningful child assent during emergency care. Participatory design methods adapted for pediatric contexts could play a valuable role in ensuring that these governance structures reflect the perspectives of affected families.

\section{Conclusion}

AI-enabled recording systems are entering pediatric emergency care with increasing technical sophistication. Their potential to improve documentation, debriefing, training, and clinical outcomes is supported by a growing body of evidence. However, the concerns raised in this paper indicate that this potential will be difficult to realize fully without more sustained attention to the perspectives of the people these systems observe. Without such reorientation, the risks are concrete: erosion of clinician and family trust, systematic behavioral bias in the data these systems generate, and growing clinical resistance that could ultimately undermine adoption itself.

We have argued that the question of who is in the room, whose perspectives inform design, whose voices shape governance, and whose experiences guide evaluation, should be as central to the development of these systems as the question of what they can technically achieve. The concerns we raise about consent, emotional impact, surveillance dynamics, and participatory governance are not merely ethical shortcomings. They are interaction design problems that shape the legitimacy, data quality, and clinical utility of the systems being built. The positions advanced here are intended as a framing for future empirical work. They require validation through multi-site qualitative studies exploring how clinicians, parents, and institutional decision-makers experience and respond to recording in pediatric emergency settings, as well as pilot evaluations of reflexive recording models and purpose-specific consent interfaces. Such studies would test whether the design orientations we propose are feasible, acceptable, and effective in practice, and would generate the evidence base needed to inform institutional policy. Advancing this reframing will require researchers who work across human-centered computing and health to build systems whose legitimacy rests not only on technical performance but on the meaningful participation of the people who are recorded.

\section*{Acknowledgments}
This work was supported by the Swiss National Science Foundation (SNSF) under Grant No.\ 10002119. The positions, arguments, and analysis presented in this paper are entirely the work of the authors. In accordance with ACM policy on the use of generative AI, the authors disclose that a large language model (Claude, Anthropic) was used as a writing assistant for language editing, including grammar correction, stylistic revision, and consistency checking. The model was not used to generate scholarly content, develop arguments, or identify or interpret sources.

\bibliographystyle{ACM-Reference-Format}
\bibliography{references}

\end{document}